\documentclass[twocolumn,showpacs,preprintnumbers,amsmath,amssymb]{revtex4}
\usepackage{graphicx}% Include figure files

\begin{document}

\title{Design of magnetic textures of ad-nanocorrals with an extra adatom}

\author{N. P. Konstantinidis$^1$ and Samir Lounis$^2$}
\affiliation{(1) Fachbereich Physik und Landesforschungszentrum OPTIMAS, Technische Universit\"at Kaiserslautern, 67663 Kaiserslautern, Germany \\
(2) Peter Gr\"unberg Institut and Institute for Advanced Simulation, Forschungszentrum J\"ulich and JARA, 52425 J\"ulich, Germany}

\date{\today}

\begin{abstract}
It is shown that in antiferromagnetic open or closed corrals of magnetic adatoms grown on surfaces, the attachment of a single extra adatom anywhere in the corral impacts on the geometrical topology of the nanosystem and generates complex magnetic structures when a magnetic field is applied or a magnetic coupling to a ferromagnetic substrate exists. The spin configuration of the corral can be tuned to a non-planar state or a planar non-collinear or ferrimagnetic state by adjusting its number of sites, the location of the extra adatom or the strength of the coupling to the ferromagnetic substrate. This shows the possibility to generate non-trivial magnetic textures with atom by atom engineering anywhere in the corral and not only at the edges.

\end{abstract}


\maketitle

In recent years artificial engineering of molecular nanomagnets, magnetic clusters and arrays 
of magnetic adatoms adsorbed on surfaces has emerged for the construction of entities with rich 
magnetic properties that can be constituents of nanospintronics devices 
\cite{Bucher91,Lau02,Manoharan00,Gambardella02,Knorr02,Silly04,Hirjibehedin06,Loth12}. 
These entities can be fabricated directly on surfaces in a bottom-up fashion with 
Scanning Tunneling Microscopy (STM) \cite{Eigler90,Ruess07,Moon09}, and STM is also used to 
directly measure their magnetic properties \cite{Meier08,Heinrich04}. The rich magnetic properties 
originate in the exchange couplings between the individual magnetic moments \cite{Lounis05,Ribeiro11,Stepanyuk05,Bergman06} and are of great 
interest for concepts like spin-transfer torque \cite{Slonczewski96,Kiselev03,Krause07} and 
spin chirality \cite{Menzel12} on the nanoscale, as well as for potential applications in quantum 
computing \cite{Leuenberger01,Troiani05,Khajetoorians11}. The STM measurements allow precise access 
to the properties of the individual magnetic moments, which when combined with tailored construction 
of the coupling between the spins provide systems where nanomagnets with desired properties can be 
synthesized \cite{Zhou10,Khajetoorians12}. In particular, arrays of a small number of 
magnetic atoms on non-magnetic metallic substrates are very promising candidates in this direction. 
This bottom-up approach is one of the main focus research areas of nanoscience where various atomic 
structures, e.g. corrals of adatoms and nanowires, are engineered atom by atom. 

It has already been 
shown theoretically using Density Functional Theory (DFT) that adatom nanochains can support 
non-collinear magnetic structures when the exchange interaction between the adatoms is 
antiferromagnetic (AFM) and a coupling to a ferromagnetic substrate exists \cite{Lounis08}. 
In the peculiar case of Mn chains on a Ni(100) substrate, the AFM coupling within the chain competes with the ferromagnetic coupling of the chain to the substrate leading to an even-odd effect, where the magnetic texture depends crucially on the parity of the 
number of atoms in the chain. The Mn chain is in an AFM configuration but if the weak ferromagnetic interaction to the substrate is switched 
on, the spins in the odd numbered chain retain their collinearity while the uncompensated moment of the chain aligns with that of the substrate. 
Even numbered chains, however, develop a more complex non-collinear ground state. This even-odd effect has been observed recently by STM in short 
Mn wires on a Ni(110) substrate \cite{Lounis13}.

For such an effect a ferromagnetic substrate is required when the exchange interaction is nearest neighbor, therefore 
strong. However, when the exchange interaction is non-nearest neighbor, therefore weaker, for example when it is mediated 
by Ruderman-Kittel-Kasuya-Yosida (RKKY) interactions \cite{RKKY1,RKKY2,RKKY3}, an external magnetic field can provide the 
necessary energy to compete with the exchange energy. Measurements with STM can be performed for both kinds of situations:
 Magnetic nanostructures on ferromagnets or on non-magnetic substrates. It is pointed out that if the magnetic anisotropy 
energy is strong (a few meV), the RKKY interactions would not be able to create non-collinear structures although there is competition between the involved magnetic interactions.

In this paper, AFM wires or corrals with a small number of magnetic adatoms are considered, where an extra 
adatom is attached anywhere along the nanostructure. It is shown that 
one more or less adatom is crucial for the magnetic properties of such nanostructures, and can generate planar or even 
non-coplanar magnetic configurations. This opens up the possibility of generating new complex magnetic structures whose 
accessibility depends on the precise location of the extra adatom, which is not necessarily attached at the 
edges of the corral, the total number of sites of the corral and the strength of the coupling to a 
ferromagnetic substrate or an external magnetic field. Besides the parity of the number of adatoms in the 
corrals, the parity of the site to which the extra atom is attached determines not only the magnetic texture of 
the whole nanostructure (open or closed), but also the magnetic behavior of the edge atoms for the open corrals.
This is of crucial importance for building logic gates made up of a few adatoms (see e.g. \cite{Khajetoorians11}). In the 
following, we will refer to wires which are not necessarily straight as open corrals.

The magnetic properties of the nanostructures are modeled with the AFM Heisenberg model (AHM). It is mainly 
considered at the classical level, but it is also shown that the classical results compare very well with 
results for typical quantum values of the magnetic moments. Previous calculations based on DFT have shown 
that the AFM Heisenberg model is reliable to predict the complex magnetic texture in such 
nanostructures \cite{Lounis08}. The AHM is considered for corrals of adatoms that form open or closed corrals, 
with an extra adatom attached to any location other than the edges. The coupling between the individual 
spins $s_i$ is $J>0$. The interaction with a ferromagnetic substrate or with an external magnetic field (in the case of a non-magnetic substrate) 
has strength $h$, and the Hamiltonian is:
\begin{equation}
H = J ( \sum_{i=1}^{N-1}{ \vec{s}_i \cdot \vec{s}_{i+1} } + \vec{s}_L \cdot \vec{s}_E ) - h ( \sum_{i=1}^N s^z_i + s_E^z )
\label{eq:Hchain}
\end{equation}
where the number of atoms of the corral is $N$, with $L=2,...,N-1$ the corral adatom away from the edges 
that couples to the extra adatom $\vec{s}_E$. For closed corrals the term $J \vec{s}_1 \cdot \vec{s}_N$ 
must be added to (\ref{eq:Hchain}) and $L$ can be any site as there are no open ends and the system is 
translationally invariant without the extra adatom. The magnetic field $\vec{h}$ lies along the $\hat{z}$ axis and tends to make 
all spins parallel, competing with the AFM configuration supported by $J$. To simplify the discussion the magnetic interaction between the extra adatom $E$ and the adatom at $L$ is taken to be $J$, but this 
is not necessarily the case. The result of varying this interaction is discussed briefly in 
the Supplemental Material. For the case of RKKY interactions, small magnetic anisotropy energies are required in order to observe 
a rich magnetic phase diagram, otherwise the adatom moments would be pinned to a collinear behavior. This is the case for Fe adatoms on a Cu(111) 
surface \cite{Khajetoorians12}. It is also demonstrated that small magnetic anisotropy energies, not considered 
in the aforementioned Hamiltonian, lead to negligible effects on the magnetic behavior of the nano-corrals (see Supp. Mat.). Thus adatoms with low magnetic anisotropy energy, e.g. $Cr$ or $Mn$, are proposed, in contrast to $Co$ or $Fe$, to be deposited on different non-magnetic substrates. $Cr$ and $Mn$ have weak magnetic anisotropy so that the AFM Heisenberg model without 
an extra anisotropy term accurately describes the magnetism of the system. 

DFT calculations predict that for Cr, Mn and Fe deposited on Cu(111) the magnetic moments are respectively 4.1, 4.3 and 3.2 $\mu_B$ \cite{Lounis06}, 
while Cr and Mn adatoms on a
Ni(001) or a Fe(001) substrate have a magnetic moment of the order of 3.5 to 4 $\mu_B$, which decreases when forming chains due to hybridization \cite{Lounis05}. Thus, typically $s_i=3/2$ or 2. It is sufficient to consider the $s_i$ as classical unit vectors, and the lowest energy configuration is 
found for any $h$ \cite{Coffey92,NPK07}. Typical quantum values are also considered \cite{NPK05,NPK09}. It has already been shown that 
for open chains with no extra spins attached the classical predictions survive for relatively low values of $s_i$ \cite{Machens13}. 
Here similar conclusions are drawn for the dependence of the magnetic properties on the magnitude of $s_i$ (see also Suppl. Mat.).

For open corrals the exact location of the extra adatom is important, due to the lack of translational symmetry of the open corral. The magnetic 
behavior depends on the parity of $N$, and also on the parity of the linking location $L$ of the extra adatom $\vec{s}_E$. For 
closed corrals the exact linking point is unimportant due to translational symmetry, and the magnetic behavior depends only 
on the parity of $N$.

\begin{figure}
\includegraphics[width=0.7\columnwidth,angle=270]{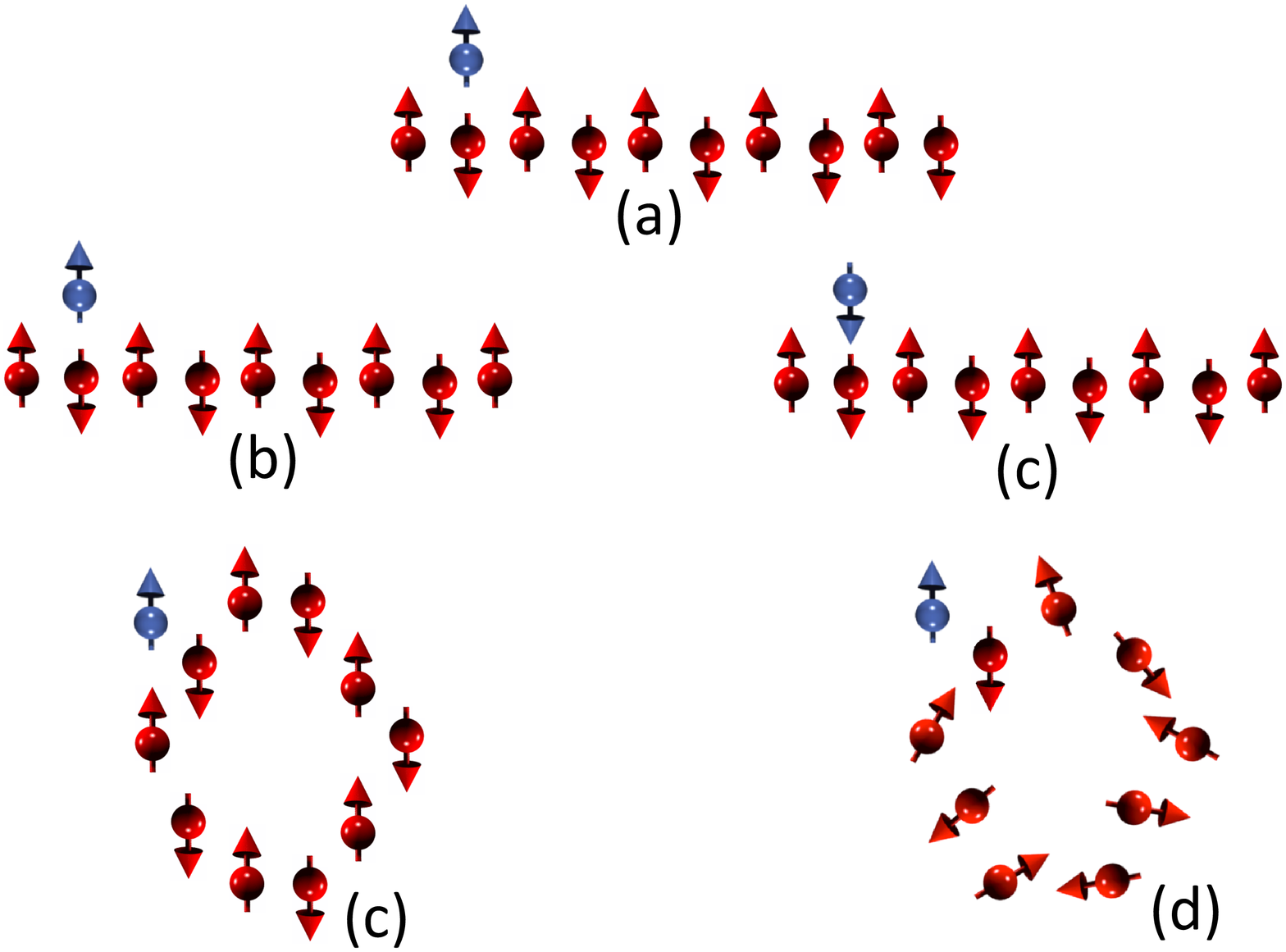}
\caption{Lowest energy spin configurations in the absence of a magnetic field (red arrows denote spins in the corral, and the blue arrow the attached spin). (a) Open even corral with a number of adatoms $N=10$ and an extra adatom at location $L=2$. The zero field magnetization is $M_{h=0}=s_i$. (b1) Open odd corral with $N=9$ and an even linking point $L=2$, $M_{h=0}=2s_i$. (b2) Open odd corral with $N=9$ and an odd linking point $L=3$, $M_{h=0}=0$. (c) Closed even corral with $N=10$, $M_{h=0}=s_i$. (d) Closed odd corral with $N=9$, $M_{h=0}=s_i$. As in (c) the value of $L$ does not make any difference due to translational invariance.}
\label{fig:corralsconfigurations}
\end{figure}

\begin{figure}
\includegraphics[width=3.5in,height=2.5in]{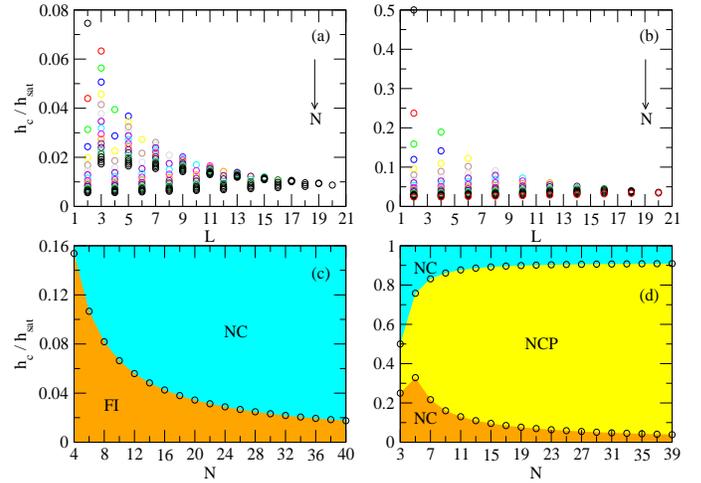}
\vspace{1pt}
\caption{Critical fields $h_c$ scaled to the saturation field $h_{sat}$ for which transitions between different configurations take place for the corrals. (a): open corrals with even number of adatoms $N$, (b): open corrals with odd $N$. The horizontal axis shows the location $L$ where the extra adatom is attached. The values of $\frac{h_c}{h_{sat}}$ are symmetric with respect to the center of the corral, so for an even corral the values for $L=2,\ldots,\frac{N}{2}$ are shown, while for an odd corral the values for $L=2,\ldots,\frac{N+1}{2}$. A single corral is represented by symbols of the same color, and the direction where $N$ increases is shown with an arrow. For fixed $L$, $\frac{h_c}{h_{sat}}$ decreases with increasing $N$. For odd corrals there is a transition only for even values of $L$. A clear difference in the behavior of $h_c$ between even and odd corrals is seen, as its values are higher for the odd corrals for the same $L$, and also monotonic with $L$, unlike even corrals. The spin configuration of an open corral can be tuned to a ferrimagnetic (FI) ($h<h_c$) or planar non-collinear (NC) state ($h>h_c$) by adjusting $N$, $L$ or $h$. (c): closed corrals with even $N$, (d): closed corrals with odd $N$. The horizontal axis is $N$. Due to translational invariance the exact value of $L$ does not affect the magnetic behavior. For even $N$ the corral changes from a FI to a NC configuration with increasing $h$, while for odd $N$ it first goes from a NC to a non-coplanar (NCP) and then to a NC configuration.}
\label{fig:figattach}
\end{figure}

{\it{Open even corrals.}} An open even corral with no extra adatom attached has nearest neighbor spins pointing in opposite 
directions in the absence of an external field. Irrespectively of the point of attachment of the extra adatom, the extra 
spin is not compensated and the total magnetization at zero field $M_{h=0} = s_i$, corresponding to a ferrimagnetic (FI) configuration 
(Fig. \ref{fig:corralsconfigurations}(a)). It takes a finite magnetic field $h_c$ to change the FI configuration, up to 
which the only gain in energy comes from the coupling to the field. The susceptibility $\chi$ is discontinuous as $h_c$ is 
crossed and for higher fields the spins are in a planar non-collinear (NC) configuration. The open even corral with an adatom 
attached to it is thus similar to an open odd chain \cite{Machens13}. $h_c$ decreases on the average with $N$ 
(Fig. \ref{fig:figattach}(a)), as $M_{h=0}$ decreases with respect to the saturation magnetization $M_{sat}=(N+1)s_i$ with $N$, 
$\frac{M_{h=0}}{M_{sat}}=\frac{1}{N+1}$. $h_c$ also depends on the parity of $L$. As $L$ changes from even to odd, the length 
of the largest sub-chain of the corral changes its parity from even (which when isolated has no FI lowest field configuration) to 
odd (which possesses a FI low field configuration), and correspondingly its zero-field magnetization from 0 to $s_i$. It is 
thus reminiscent of an isolated even or odd chain respectively, especially since the corral feels the influence of the extra 
adatom more strongly around its linking site $L$, and its influence gradually weakens when going away 
from it. Therefore the length of the largest odd sub-chain fluctuates significantly with $L$ when the linking point is away 
from the middle, generating the pronounced non-monotonic dependence of $h_c$ on the parity of $L$ (Fig. \ref{fig:figattach}(a)), 
due to the monotonic dependence of the critical field on the length for an isolated (odd) chain. When 
the extra adatom approaches the center of the chain the length of the largest odd sub-chain does not change significantly and $h_c$ tends to a constant.

\begin{figure}
\includegraphics[width=3.1in,height=2.1in]{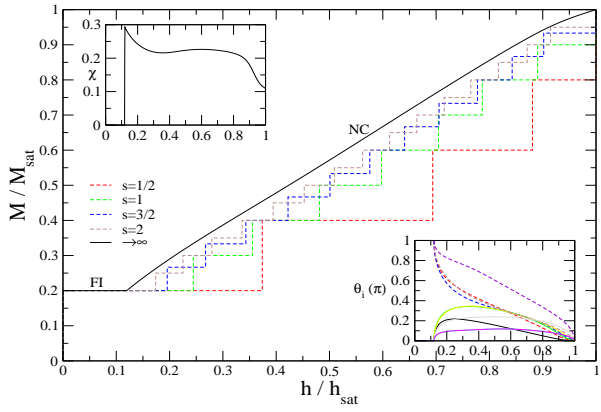}
\caption{Reduced magnetization $M/M_{sat}$ as a function of $h/h_{sat}$ for an open corral with $N=9$ and an extra adatom linked at site $L=8$. The classical result corresponds to a ferrimagnetic (FI) state for lower fields, and a planar non-collinear (NC) state for higher fields. The quantum mechanical magnetization tends to the classical curve as $s$ increases from 1/2 towards the typical values 3/2 and 2. The classical susceptibility $\chi$ (upper left inset) has a discontinuity at $h_c=0.120h_{sat}$. The polar angles in units of $\pi$ (lower right inset) are on the average turning towards the field. For smaller fields above the discontinuity spins that originally point along the field direction turn away from the field to increase the exchange energy. Nearest neighbor azimuthal angles differ by $\pi$, with the two groups of spins having the same azimuthal angle indicated by solid (spins at odd sites and extra spin) and dashed (spins at even sites) lines respectively.}
\label{fig:figclassquant}
\end{figure}

{\it{Open odd corrals.}} An open corral with an odd number of atoms has a single uncompensated spin. When an extra adatom is 
attached $M_{h=0}$ depends on the parity of the linking point $L$. If $L$ is even the extra spin is parallel to the uncompensated 
spin of the open corral and $M_{h=0} = 2s_i$ (Fig. \ref{fig:corralsconfigurations}(b1)). In contrast, if $L$ is odd the extra 
spin balances out the uncompensated spin of the corral and $M_{h=0} = 0$ (Fig. \ref{fig:corralsconfigurations}(b2)). For even $L$ 
the lowest energy configuration is FI, similarly to an open even corral with an extra adatom. However there are now two 
uncompensated spins, consequently $h_c$ is bigger (Fig. \ref{fig:figattach}(b)). For the same reason $h_c$ is also bigger compared 
to an open even corral, which also has only a single uncompensated spin. For higher fields the lowest energy configuration is also 
NC as in the open even corral case (Fig. \ref{fig:figclassquant}). The change between the FI and the NC configuration also 
generates a discontinuity in the magnetic susceptibility $\chi$ (upper left inset of Fig. \ref{fig:figclassquant}), with nearest-neighbor 
spins pointing in opposite directions in the azimuthal plane for higher fields (lower right inset of Fig. \ref{fig:figclassquant}). On the other hand, 
when $L$ is odd, the lowest energy configuration is AFM for $h=0$ and $M$ immediately responds to an external field with no 
susceptibility discontinuity (Fig. \ref{fig:figattach}(b)), leading to a NC configuration, similarly to an even corral with no extra 
adatoms. Consequently there are changes in the lowest energy configuration as a function of $h$ only for even $L$, where the size of 
the largest odd sub-chain of the isolated corral decreases with $L$. Thus the critical field for increasing even $L$ increases as the zero-field 
sub-chain magnetization increases with respect to the sub-chains saturation magnetization, in contrast to the non-monotonic result of the open even corrals. This shows 
the importance of the exact location of attachment of the extra adatom for the magnetic configuration of the corral, leading to an 
even-odd linking effect for open odd corrals, which is absent for open even corrals. These clear-cut differences between open-even and open-odd corrals 
provide the opportunity for a variety of magnetic structures as the length of the corral is varied and the extra adatom is moved along the corral with STM.

Up to now, classical spins have been considered in Hamiltonian (\ref{eq:Hchain}). However, as mentioned earlier, typical values for the corral spins 
are estimated to be $s_i=3/2$ or $2$ and magnetization curves for $s_i \leq 2$ are shown in 
Fig. \ref{fig:figclassquant}, along with the classical result. The quantum mechanical magnetization approaches the classical 
curve with increasing $s_i$, quite close already for $s_i=2$. Therefore the qualitative features of the classical calculation 
are expected to survive for the actual $s_i$ values, in accordance with a similar calculation for open chains \cite{Machens13}.

{\it{Closed even corrals.}} In a closed corral, in contrast to open corrals, due to translational invariance of the isolated corral the exact linking 
point $L$ of the extra adatom does not affect the magnetic behavior. For an isolated even corral there is no frustration originating in 
the closed boundary conditions, and nearest neighbor spins are antiparallel with no net magnetization in the absence of a field 
(Fig. \ref{fig:corralsconfigurations}(c)). The extra adatom provides an uncompensated spin and $M_{h=0}=s_i$, and it takes a finite 
magnetic field to destroy the FI configuration and generate a NC one, similarly to the previous cases. In the NC 
state spins symmetrically placed with respect to the linking adatom $\vec{s}_L$ point in the same direction. The dependence 
of $h_c$ on $N$ is shown in Fig. \ref{fig:figattach}(c). As was the case before, $M_{h=0}$ decreases with respect to $M_{sat}$ as $1/(N+1)$, thus $h_c$ decreases with $N$.

\begin{figure}
\begin{minipage}{0.33\textwidth}
\includegraphics[width=\textwidth,height=1.2in]{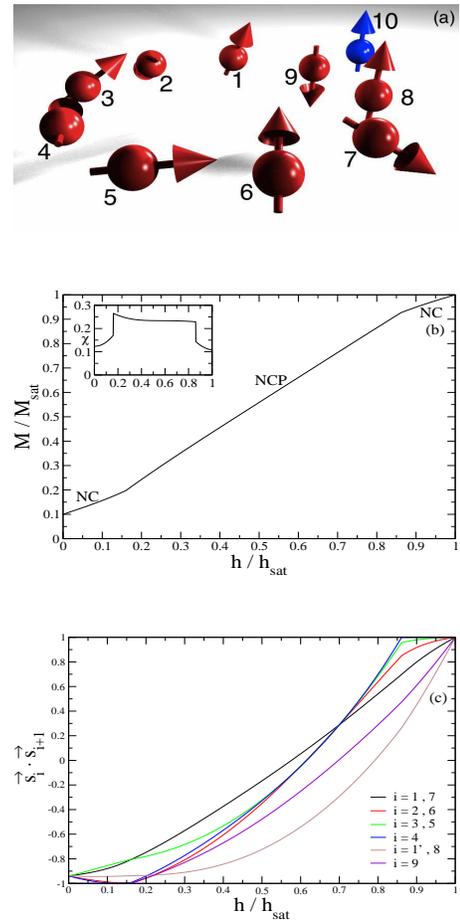}
\vspace{10pt}
\end{minipage}
\begin{minipage}{0.33\textwidth}
\includegraphics[width=\textwidth,height=1.5in]{Figure4b.eps}
\vspace{10pt}
\end{minipage}
\begin{minipage}{0.33\textwidth}
\includegraphics[width=\textwidth,height=1.5in]{Figure4c.eps}
\end{minipage}
\caption{A closed corral with $N=9$ and an extra adatom. (a) Non-coplanar (NCP) configuration of spins for $h=0.171h_{sat}$. (b) Reduced magnetization $M/M_{sat}$ as a function of $h/h_{sat}$. The classical susceptibility (inset) has two discontinuities at $h/h_{sat}=0.160$ and $0.861$, which divide the planar non-collinear (NC) configurations from the NCP. (c) Nearest neighbor correlation functions $\vec{s}_i \cdot \vec{s}_{i+1}$ as a function of $h/h_{sat}$, where $1'$ indicates correlation $\vec{s}_1 \cdot \vec{s}_9$. They are symmetric with respect to the linking adatom $\vec{s}_L$ of the extra spin $\vec{s}_E$ (here $L=9$ and $E=10$). For low fields there are correlations that get stronger with $h$.}
\label{fig:N=9ring}
\end{figure}

{\it{Closed odd corrals.}} For odd $N$ the periodic boundary condition introduces a frustrated configuration even for an isolated 
closed corral with no extra adatoms. It is not possible for nearest-neighbor spins to be antiparallel even when $h=0$, however the 
net magnetization is still zero. The extra adatom provides an uncompensated spin and $M_{h=0}=s_i$, but the configuration now is NC, 
unlike the FI configurations found before (Fig. \ref{fig:corralsconfigurations}(d)). This configuration is susceptible to an 
infinitesimal magnetic field, again unlike all previous cases with $M_{h=0} \neq 0$ (Fig. \ref{fig:N=9ring}(b)). In it the extra and the linking adatom 
are always antiparallel, with the extra adatom always pointing along the direction of the field. Spins which are symmetrically 
placed with respect to the linking adatom $\vec{s}_L$ have the same polar angle (this is true for all magnetic fields up to saturation), 
but they point in opposite directions in the azimuthal plane (see also Supp. Mat.). In addition, nearest neighbor correlations between 
specific pairs of spins increase with $h$, and eventually 
these spins become antiparallel (Fig. \ref{fig:N=9ring}(c)). These pairs are arranged in a ``dimer'' type of configuration, with every 
second bond increasing or decreasing in correlation strength. From the ones which are getting stronger with $h$, they increase in 
correlation strength as the linking point $L$ of the extra adatom is approached. This is counter-intuitive and points to the importance 
of the single extra spin for the change of the magnetic properties of the whole corral.

With increasing field, a susceptibility discontinuity leads for the first time to a non-coplanar (NCP) lowest spin configuration 
(Fig. \ref{fig:N=9ring}(b)), shown in Fig. \ref{fig:N=9ring}(a) for $h=0.171h_{sat}$. The nearest neighbor correlations now decrease only 
with the magnetic field (Fig. \ref{fig:N=9ring}(c)), even though there are initially polar angles that do not decrease with $h$, 
due to the competition of exchange and magnetic energy (see also Supp. Mat.). For higher fields close to saturation a second 
susceptibility discontinuity leads to a NC phase.
Now spins symmetrically placed with respect to the linking adatom $\vec{s}_L$ point in the same direction. The transition fields and 
the range of existence of the different phases as function of $N$ are shown in Fig. \ref{fig:figattach}(d). The odd closed corral 
with the extra adatom combines the competition between exchange and magnetic energy with the frustration introduced by the closed 
boundary conditions, supporting a NCP configuration not found in any of the other cases. In addition, it supports two susceptibility 
discontinuities in its magnetization curve. Therefore its magnetic behavior is in direct contrast with its even counterpart and the 
open corrals as well.

To conclude, the magnetic properties of corrals (open or closed) with an extra adatom attached are different from their counterparts 
without the extra adatom. The extra spin changes the geometrical topology of the nanosystem, which has an impact on its magnetic behavior as a whole. It 
can tune the lowest energy configuration of the spins according to the parity of the 
corral, the parity of the linking point of the extra spin and the presence or not of periodic boundary conditions. This leads to a plethora of 
configurations that can also be three-dimensional, and can be accessed and of use experimentally.

N. P. K. acknowledges support by the Graduate School of Excellence MAINZ.
S. L. acknowledges the support of the HGF-YIG Program No. VH-NG-717
(Functional nanoscale structure and probe simulation
laboratory-Funsilab).

\section{Supplemental Material}

\subsection{Extra Spin Linked with Varying Strength}
\begin{figure}
\includegraphics[width=3.1in,height=2.1in]{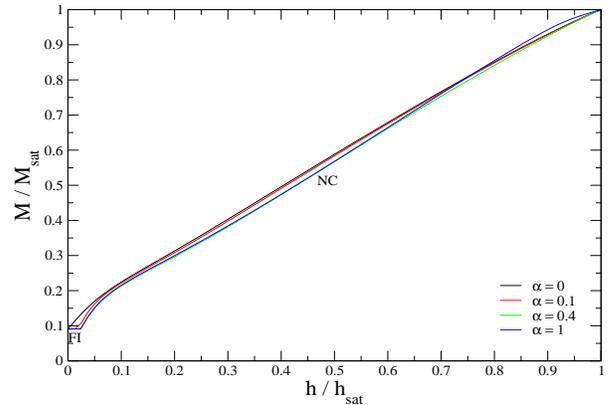}
\caption{Reduced magnetization $M/M_{sat}$ as a function of $h/h_{sat}$ for an open corral with $N=10$ and an extra spin linked at site $L=9$, where the strength $J$ of the $\vec{s}_L \cdot \vec{s}_E$ bond in Hamiltonian (1) of the main text is scaled with $\alpha$. The ferrimagnetic part of the magnetization curve (low fields) is robust, even for small values of $\alpha$.}
\label{fig:linkstrength}
\end{figure}

The robustness of the influence of the extra adatom on the isolated corral can be shown by varying the strength of the coupling $\vec{s}_L \cdot \vec{s}_E$, which is equal to $J$ in Hamiltonian (1) of the main text. If this coupling is scaled with $\alpha$, then the magnetization of an $N=10$ open corral with an extra adatom attached at site $L=9$ is shown in Fig. \ref{fig:linkstrength}. Even for small values of $\alpha$ the ferrimagnetic  state, absent when $\alpha=0$, appears for small magnetic fields.

\subsection{Comparison of Classical and Quantum Results for Moderate $s_i$}
\begin{figure}
\vspace{10pt}
\includegraphics[width=3.1in,height=2.1in,angle=0]{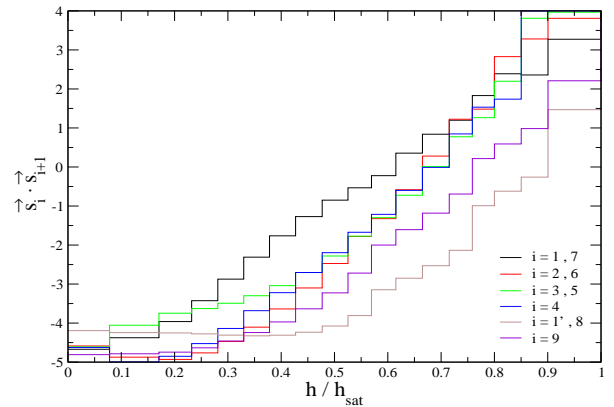}
\caption{Nearest neighbor correlation functions $\vec{s}_i \cdot \vec{s}_{i+1}$ for a $N=9$ closed corral with an extra adatom attached for $s_i=2$ as a function of $h/h_{sat}$, where $1'$ indicates correlation $\vec{s}_1 \cdot \vec{s}_9$. The results are similar to the classical results (compare Fig. 4(c) of the main text).}
\label{fig:correlationsN=9ringplus1s=2}
\end{figure}

In Ref. [37] of the main text it has been shown that for open isolated corrals without an attached adatom the classical result describes accurately the physics up to relatively small quantum numbers, $s_i \gtrsim 3/2$. As shown in Fig. 3 of the main text this is similar for the case of the corrals with an extra adatom, as the classical magnetization is closely approached already for such $s_i$ values. In Fig. 4(c) of the main text the classical nearest neighbor correlation functions are plotted as a function of $h/h_{sat}$. The corresponding plot for $s_i=2$ is shown in Fig. \ref{fig:correlationsN=9ringplus1s=2}, where by comparison it is seen that the classical result again already provides a very good description.

\subsection{Polar Angles for the $N=9$ Closed Corral}
\begin{figure}
\vspace{20pt}
\includegraphics[width=3.1in,height=2.1in]{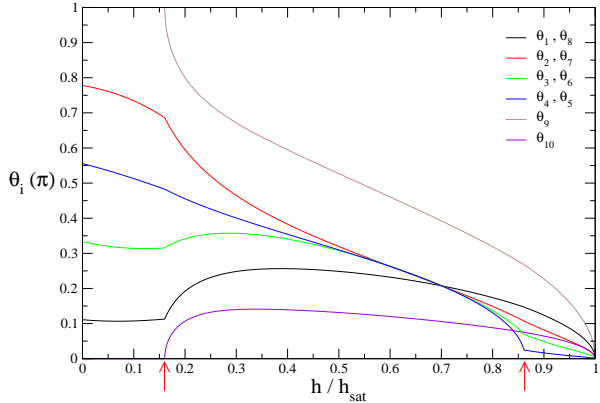}
\caption{Polar angles in units of $\pi$ as a function of $h/h_{sat}$ for the $N=9$ closed corral with a single spin attached. The two susceptibility discontinuities at $h/h_{sat}=0.160$ and $0.861$ (shown with red arrows) separate the three different lowest energy configurations, which change from NC to NPC and then to NC with increasing field.}
\label{fig:N=9corral}
\end{figure}

The polar angles for the closed corral with $N=9$ and an extra adatom are shown in Fig. \ref{fig:N=9corral}. Polar angles symmetrically placed with respect to the linking spin $\vec{s}_L$ are equal. The three different regimes for the lowest energy configuration as a function of $h/h_{sat}$ are clearly seen, separated by two susceptibility discontinuities. The dependence of the $\theta_i$ on $h$ is in general non-monotonic.

\subsection{Magnetic Anisotropy}
\begin{figure}
\includegraphics[width=3.1in,height=2.1in]{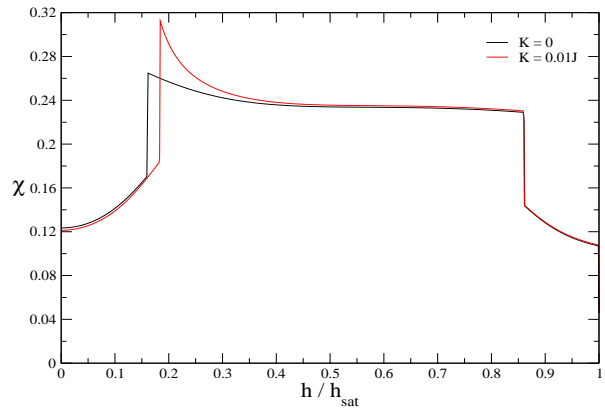}
\caption{Magnetic susceptibility $\chi$ for a $N=9$ closed corral with an extra adatom, without and with magnetic anisotropy $K$, with the easy axis parallel to the direction of the magnetic field. A small magnetic anisotropy $K=0.01J$ does not change significantly the magnetic response, which has two susceptibility discontinuities at $h/h_{sat}=0.160$ and $0.861$ for $K=0$, and at $h/h_{sat}=0.184$ and $0.861$ for $K=0.1J$.}
\label{fig:anisotropy}
\end{figure}

The influence of easy-axis magnetic anisotropy can be seen in Fig. \ref{fig:anisotropy}, where a term $-K[\sum_{i=1}^N (s_i^z)^2+(s_E^z)^2]$ is added in Hamiltonian (1) of the main text, with the easy axis parallel to the magnetic field axis. For a $N=9$ closed corral with an extra adatom and weak anisotropy $K=0.01J$ the magnetic response does not change significantly.

\bibliography{corrals}

\end{document}